\documentclass[9pt]{article}

\usepackage{spconfa4,amsmath,graphicx}
\usepackage{graphicx}%
\usepackage{multirow}%
\usepackage{amsmath,amssymb,amsfonts}%
\usepackage{amsthm}%
\usepackage{mathrsfs}%
\usepackage[figuresright]{rotating}%
\usepackage[title]{appendix}%
\usepackage{xcolor}%
\usepackage{textcomp}%
\usepackage{manyfoot}%
\usepackage{booktabs}%
\usepackage{algorithm}%
\usepackage{algorithmicx}%
\usepackage{algpseudocode}%
\usepackage{program}%
\usepackage{listings}%
\usepackage{cite}%

\title{Mixture-of-Experts Framework for Field-of-View Enhanced Signal-Dependent Binauralization of moving talkers}
\name{%
  \begin{tabular}{@{}c@{}}
  Manan Mittal$^{1,3}$, Thomas Deppisch$^{2,3}$, Joseph Forrer$^3$, Chris Le Sueur$^3$, \\
  Zamir Ben-Hur$^3$, David Lou Alon$^3$, Daniel D.E. Wong$^3$
  \end{tabular}
}
\address{Stony Brook University$^1$, Chalmers University of Technology$^2$, Reality Labs Research, Meta$^3$}

\begin{document}
\topmargin=0mm

\maketitle

\begin{abstract}
We propose a novel mixture of experts framework for field-of-view enhancement in binaural signal matching. Our approach enables dynamic spatial audio rendering that adapts to source motion, allowing users to emphasize or suppress sounds from selected directions while preserving natural binaural cues. Unlike traditional methods that rely on explicit direction-of-arrival estimation, our signal-dependent framework combines multiple binaural filters in an online manner using implicit localization. This allows for real-time tracking and enhancement of moving sound sources, supporting applications such as speech focus, noise reduction, and world-locked audio in augmented and virtual reality. The method is agnostic to array geometry and offers a flexible solution for spatial audio capture and playback in next-generation consumer audio devices.
\end{abstract}
\section{Introduction}
\label{sec:intro}
Consumer audio capture devices are increasingly designed as wearable technologies. Among these, headworn microphone arrays have gained significant attention for capturing sound fields and enabling binaural rendering. A key use case arises when the user wishes to re-experience the recording in a way that matches how it originally sounded. This places importance on downstream processing methods that preserve the auditory cues present at the time of capture~\cite{Rafaely2022}. The process of filtering and summing the microphone array signals to reproduce the binaural cues at the left and right ears of the user has been referred to as end-to-end magnitude least squares (eMagLS)~\cite{Deppisch2021a} or binaural signal matching (BSM)~\cite{Madmoni2024a}. An alternative method to render binaural signals relies on direction-of-arrival (DOA) estimation and beamforming to extract direct signal components and renders direct and reverberant sound field components separately~\cite{Politis2018}. 

\begin{figure}
    \label{fig:system_diagram}
    \centering
    \centerline{\includegraphics[width=1.1\linewidth, height=4cm]{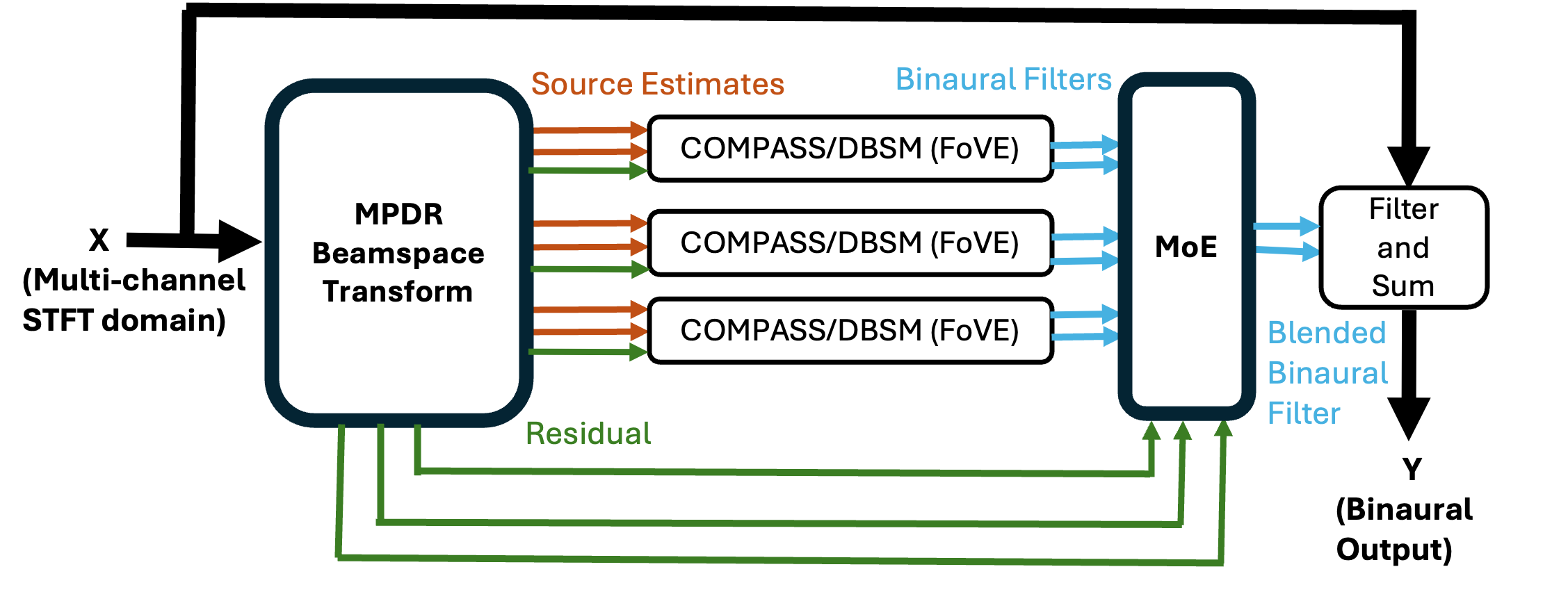}}
    \caption{The proposed binauralization method. Each directional output of the beamspace transform is rendered using signal-dependent binaural signal matching. Each candidate direction is deemed an "expert" and the methods seeks to adaptively blend the estimates from different experts.}
    \vspace{-10mm}
\end{figure}

It is desirable to give the user additional control over properties of the rendered audio. Such control may take the form of speech enhancement, noise reduction, support for world-locked audio where playback adapts to user movement, or directional enhancement where sounds from selected directions or from the user’s gaze are emphasized \cite{schultz2010acoustical, kronlachner2014spatial, 9768239}. This paper develops a signal-dependent framework for the latter task, which we refer to as field-of-view enhancement (FoVE)~\cite{Fernandez2024a}. Moreover, we propose a mixture of experts algorithm that is able to combine the estimates of numerous signal-dependent binaural signal matching filters in an online manner with implicit localization instead of relying on traditional direction of arrival estimators. This allows the model to track a continuous talker without assumptions of stationarity. Fig.~\ref{fig:system_diagram} depicts the system diagram for the proposed method. 
The objective of FoVE is to maintain the spatial structure of the recorded sound field while allowing the user to manipulate the directional response of the binaural rendering filter. In practice, this means enabling the user to increase or decrease the relative energy from chosen directions in a perceptually consistent manner. This produces emphasis of sources in the selected region and suppression of sources outside it, while preserving binaural cues and spatial consistency across directions. Importantly, this is not restricted to speech sources but applies to more general sound fields. The work is presented in the context of smartglasses but does not assume a specific array geometry, and the framework extends to arbitrary microphone arrays. 
\section{Signal Model}
Consider a microphone array with $N_m$ microphones used to capture an acoustic scene. We assume that the recorded sound field can be expressed as a superposition of signals arriving from $N_s$ distinct directions. In the short-time Fourier transform (STFT) domain, the signal observed at the array, at time index $t$ and frequency index $f$, is written as

\begin{equation}
    \mathbf{x}[t,f] = \mathbf{A}[f] \, \mathbf{s}[t,f] + \mathbf{v}[t,f],
\end{equation}
where $\mathbf{x}[t,f] \in \mathbb{C}^{N_m}$ is the vector of microphone signals, $\mathbf{A}[f] \in \mathbb{C}^{N_m \times N_s}$ is the acoustic transfer matrix, $\mathbf{s}[t,f] \in \mathbb{C}^{N_s}$ is the vector of source signals, and $\mathbf{v}[t,f] \in \mathbb{C}^{N_m}$ represents additive and/or diffuse noise. The dependence on time and frequency is omitted for brevity in the rest of the paper.

The second-order statistics of the signals are described by the covariance matrices
\begin{equation}
\begin{split}    
    \mathbf{R}_x &= \mathbb{E}\{\mathbf{x}\mathbf{x}^H\} \in \mathbb{C}^{N_m \times N_m}, \qquad \\
    \mathbf{R}_s &= \mathbb{E}\{\mathbf{s}\mathbf{s}^H\} \in \mathbb{C}^{N_s \times N_s}.
\end{split}
\end{equation}

In the next section, we review both signal-dependent and independent variants of binaural signal matching. At high frequencies, the magnitude-LS variant replaces the complex LS objective while using the same $\hat{\mathbf{R}}_{s}$ construction~\cite{Schorkhuber2018,Deppisch2021a,Madmoni2024a}.

\section{Binaural Signal Matching}\label{sec4}
\subsection{Signal-Independent Binaural Signal Matching}
Signal-independent BSM aims to design a linear filter that maps the microphone array signals to binaural signals at the user’s ears~\cite{Deppisch2021a,Madmoni2024a}. The design does not depend on a specific source signal but instead assumes a diffuse sound field. This corresponds to energy being uniformly distributed across all directions of arrival. The least-squares optimal binaural filter weights $\mathbf{c}^{\text{BSM}} \in \mathbb{C}^{2 \times N_m}$ solve, 
\begin{equation}
    \min_{\mathbf{c}} \; \left\| \mathbf{c}\mathbf{A} - \mathbf{H}^T \right\|_F^2
\end{equation}
where $\mathbf{H} \in \mathbb{C}^{N_s\times 2}$ denotes the head-related transfer functions (HRTFs) for the left and right ears. Under the diffuse assumption, the least-squares optimal signal-independent BSM solution takes the form
\begin{equation}
    \mathbf{c}^{\text{BSM}} = \, \mathbf{H}^T \,\mathbf{A}^H \big( \epsilon \mathbf{I} + \mathbf{A}\mathbf{A}^H \big)^{-1} .
\end{equation}
Here, $\epsilon$ is a small regularization parameter to ensure numerical stability.
\vspace{-5mm}
\subsection{Signal-Dependent Binaural Signal Matching}
In this section, we introduce signal-dependent approaches to binaural signal matching. In particular, we introduce the proposed mixture of experts approach, that may be viewed as an extension of the two other methods.
\subsubsection{COMPASS-BSM (COM)}
COMPASS is a parametric method used to render spatial audio scenes~\cite{Politis2018}. The directions of arrival of sources are estimated first. Given those directions, the received soundfield is decomposed into direct and residual parts, where the direct part is obtained as the linearly constrained minimum variance (LCMV)~\cite{van2002optimum} estimate 
\begin{equation}
\hat{\mathbf{s}}_{d} = \mathbf{W}_{d}\,\mathbf{X}, 
\qquad 
\mathbf{W}_{d} = \big(\mathbf{A}_{d}^{H}\mathbf{R}_{x}^{-1}\mathbf{A}_{d}\big)^{-1}\mathbf{A}_{d}^{H}\mathbf{R}_{x}^{-1},
\end{equation}
\noindent
where $\mathbf{A}_{d}$ stacks steering vectors of the direct component.
The binaural estimate of the direct component is rendered directly through the HRTF for each corresponding source direction stacked in $\mathbf{H}_{d}$ for the two ears,
\begin{equation}\label{direct_COM}
\hat{\mathbf{p}}^{\,d} = \mathbf{H}_{d}^{T}\hat{\mathbf{s}}_{d}.
\end{equation}
The direct component is backprojected and subtracted from the received signal to obtain the residual
\begin{equation}
\hat{\mathbf{X}}_{r} = \big(\mathbf{I}-\mathbf{A}_{d}\mathbf{W}_{d}\big)\mathbf{X}.
\end{equation}
This residual component is rendered via standard BSM weights $\mathbf{c}^{\text{BSM}}$. Therefore, the binaural residual estimate is, 
\begin{equation}
\hat{\mathbf{p}}^{\,r} = \big(\mathbf{c}^{\text{BSM}}\big)\hat{\mathbf{X}}_{r}.
\end{equation}
The final binaural output $\hat{\mathbf{p}}$ is obtained by summing direct and residual components and can be described via an equivalent single-stage filter:
\begin{equation}
\begin{split}
\hat{\mathbf{p}} &=\hat{\mathbf{p}}^{\,d}+\hat{\mathbf{p}}^{\,r}
= \big(\mathbf{c}^{\text{COM}}\big)\mathbf{X},
\qquad \\
\mathbf{c}^{\text{COM}} &= \mathbf{c}^{\text{BSM}}\big(\mathbf{I}-\mathbf{A}_{d}\mathbf{W}_{d}\big)+\mathbf{H}_{d}^{T}\mathbf{W}_{d}.
\end{split}
\end{equation}
\vspace{-3mm}
\subsubsection{Directional BSM (d-BSM)}
Signal-dependent BSM (d-BSM) utilizes a source covariance that embeds the estimated direct component, which is estimated as in COMPASS, and a diffuse reverberant component\cite{berger2024performance}:
\begin{equation}
\hat{\mathbf{R}}_{s}=\hat{\mathbf{R}}_{s_{d}}+\hat{\sigma}_{r}^{2}\mathbf{I},
\qquad
\hat{\mathbf{R}}_{s_{d}}=\mathbb{E}\{\hat{\mathbf{s}}_{d}\hat{\mathbf{s}}_{d}^{H}\},
\end{equation}
where $\hat{\sigma}_{r}^{2}$ is obtained from the residual consistent with the diffuse assumption in the paper.
Let $\mathbf{A}=[\,\mathbf{A}_{d}\ \mathbf{A}_{r}\,]$ stack the direct and reverberant steering matrices and  The signal-dependent BSM filter is
\begin{equation}
\mathbf{c}^{\text{d-BSM}}=\mathbf{H}^T\,\ \hat{\mathbf{R}}_{s}\, \mathbf{A}^{H}\big(\mathbf{A}\hat{\mathbf{R}}_{s}\mathbf{A}^{H}+\mathbf{R}_{n}\big)^{-1},
\end{equation}
with $\mathbf{R}_{n}$ the noise covariance. 
\vspace{-4mm}
\subsection{Proposed Method}
We view the binaural rendering problem as a mixture of experts \cite{shalev2012online, 720534}, where each expert corresponds to a binaural filter designed under the assumption of a single source from candidate directions. Let the set of candidate directions be indexed by $q=1,\ldots,Q$. For each direction $q$, we compute a signal dependent binaural filter $\mathbf{c}_{q}[f]$, assuming the sound source originates from that direction, and apply it to the microphone signals $\mathbf{X}[t,f]$ to obtain
\begin{equation}
    \hat{\mathbf{p}}_{q}[t,f] = \mathbf{c}_{q}^{H}[f] \, \mathbf{X}[t,f].
\end{equation}
\noindent
The final output is a convex combination of the experts,
\begin{equation}
    \hat{\mathbf{p}}[t,f] = \sum_{q=1}^{Q} \alpha_{q}[t,f] \, \hat{\mathbf{p}}_{q}[t,f],
\end{equation}
with nonnegative weights satisfying $\sum_{q}\alpha_{q}[t,f] = 1$ for each frequency $f$.  

Following the online convex optimization framework, we update the weights with exponential weighting. Let $L_{q}[t,f]$ represent the cumulative loss of expert $q$ at time $t$ and frequency $f$. The blend weights are then given by
\begin{equation}
    \alpha_{q}[t,f] = \frac{\exp\!\left(-\eta L_{q}[t,f]\right)}{\sum_{j=1}^{Q}\exp\!\left(-\eta L_{j}[t,f]\right)},
\end{equation}
where $\eta > 0$ is a learning rate. This scheme ensures that experts with lower cumulative loss contribute more strongly to the final output while preserving smooth adaptation. Additionally, it ensures that our method is time-adaptive despite using linear-time invariant filters in its construction.
The instantaneous loss $\ell_{q}[t,f]$ is defined as the squared norm of the residual, summed over all microphones $N_m$. This loss is accumulated over time into the cumulative loss $L_{q}[t,f]$, which is updated recursively. This yields an expression for the the binaural filters produced by the proposed method. 
\begin{align}
\mathbf{r}_{q}[t,f] &= \mathbf{X}[t,f] - \mathbf{A}_{q}[f] \, \hat{s}_{q}[t,f], \forall q \in {1,.., Q}\\
\ell_{q}[t,f] &= \sum^{N_m}\|\mathbf{r}_{q}[t,f]\|^{2}, \\
L_{q}[t,f] &= L_{q}[t-1, f] + \ell_{q}[t,f], \\
\mathbf{c}^\text{MoE}[t, f] &= \sum_{q=1}^{Q} \alpha_{q}[t,f] \, \mathbf{c}_{q}[t, f], \\
\hat{\mathbf{p}}[t,f] &= \big(\mathbf{c}^{\text{MoE}}[t,f]\big)^{H} \, \mathbf{X}[t,f].
\end{align}
Intuitively, the method behaves like an implicit DOA estimator that relies on the assumption that the beamformer with the lowest residual likely contains the direct source.
\noindent
\subsubsection{Regret Bound}
The exponential weighting scheme enjoys a standard regret guarantee in the online convex optimization setting. Define the regret after $T$ time frames as
\begin{equation}
    R_{T} = \sum_{t=1}^{T} \sum_{f} \ell\big(\hat{\mathbf{p}}[t,f]\big) 
    - \min_{q} \sum_{t=1}^{T} \sum_{f} \ell\big(\hat{\mathbf{p}}_{q}[t,f]\big),
\end{equation}
that is, the cumulative loss of the mixture compared to the best single expert in hindsight. With $\ell(\cdot)$ convex and bounded, the regret of the exponential weighting algorithm satisfies
\begin{equation}
    R_{T} = \mathcal{O}\big(\sqrt{T \log Q}\big).
\end{equation}
\noindent
This guarantee implies that, asymptotically, the mixture performs nearly as well as the best expert chosen with full hindsight, while retaining the flexibility to adapt to time-varying acoustic scenes. Note that the average regret $R_{T}/T$ goes to zero as time approaches infinity~\cite{singer2002universal}.
\section{Field of View Enhancement}
We now describe two control strategies for field of view (FoV) enhancement. Each strategy modifies the binaural signal matching (BSM) formulation to emphasize directions within a user-selected field of view while attenuating those outside it. Both signal-independent and signal-dependent variants are presented.

\subsection{Gain Control}
In gain control, we apply multiplicative gains to the HRTFs~\cite{Fernandez2024a}. For each direction $q$, the modified HRTF is
\begin{equation}
    \tilde{\mathbf{H}}_{q} =
    \begin{cases}
        \mathbf{H}_{q}, & q \in \text{FoV}, \\
        (1-\gamma)\,\mathbf{H}_{q}, & q \notin \text{FoV},
    \end{cases}
\end{equation}
where $\gamma \in [0,1]$ is a user-selected gain parameter. Note, that for COMPASS we can add gain $\geq 1$ in the FoV directions that are rendered directly through the HRTF as shown in~\eqref{direct_COM}.


\subsection{Distortion Control}
In distortion control, we penalize deviations in the BSM matching differently for FoV and non-FoV directions. Define the diagonal distortion weighting matrix
\begin{equation}
    \mathbf{D} = \operatorname{diag}(w_{1},\ldots,w_{N_{s}}),
    \qquad
    w_{q} =
    \begin{cases}
        1, & q \in \text{FoV}, \\
        1-\delta, & q \notin \text{FoV},
    \end{cases}
\end{equation}
with $\delta \in [0,1]$.  
The weighted least-squares BSM formulation can be written compactly for as
\begin{equation}
    \min_{\mathbf{c}} \; \big(\mathbf{c}\mathbf{A} - \mathbf{H}\big)\mathbf{D} \big(\mathbf{c}\mathbf{A} - \mathbf{H}\big)^{H},
\end{equation}
In this form, FoV directions are matched exactly while non-FoV directions are permitted distortion proportional to $1-\delta$.

In d-BSM, the distortion weights enter the covariance matrices. The weighted source covariance is
\begin{equation}
    \tilde{\mathbf{R}}_{s} = \mathbf{D}^{1/2}\mathbf{R}_{s}\mathbf{D}^{1/2},
\end{equation}
and the corresponding d-BSM filter becomes
\begin{equation}
    \mathbf{c} = \mathbf{H}^T\,\tilde{\mathbf{R}}_{s}\,\mathbf{A}^{H}\big(\mathbf{A}\,\tilde{\mathbf{R}}_{s}\,\mathbf{A}^{H} + \mathbf{R}_{n}\big)^{-1}.
\end{equation}
\subsection{Proposed FoVE via Mixture of Experts}
\begin{figure}
    \centering
    \centerline{\includegraphics[width=\linewidth]{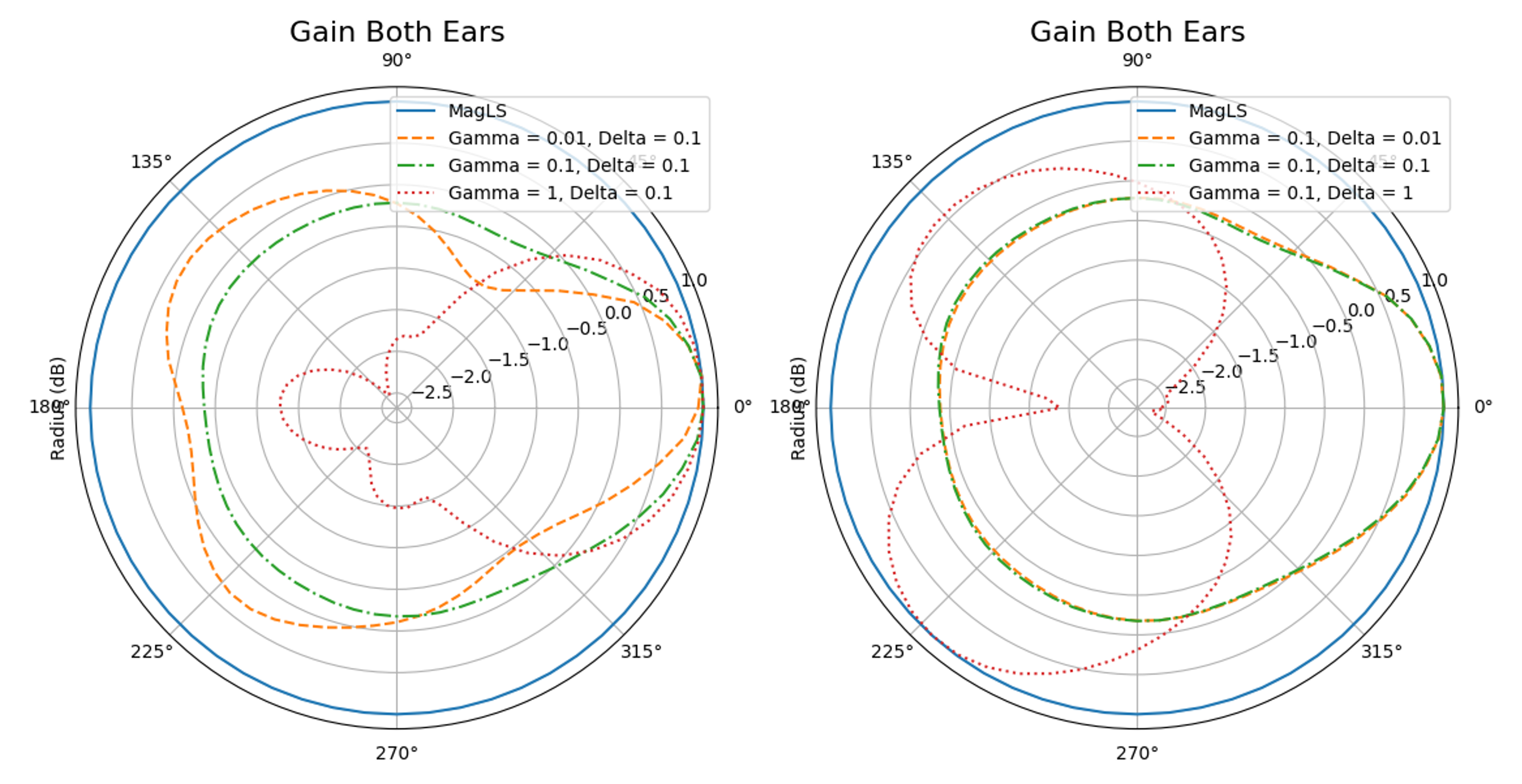}}
    \caption{The direction dependent gain from field of view in the frontal direction. The plot on left shows the effect of varying $\gamma$ with a fixed value of $\delta$ while the second plot shows the effect of varying $\delta$ with fixed value of $\gamma$.}
    \label{fig:gain}
    \vspace{-4mm}
\end{figure}

The mixture of experts framework described earlier can be directly extended to field-of-view enhancement simply by constructing the expert filters using any of the FoV-aware versions of BSM, COMPASS-BSM, or d-BSM. As shown in Fig.~\ref{fig:gain}, the resulting mixture successfully places emphasis on user-selected FoV regions.
\vspace{-5.2mm}
\section{Results}
\vspace{-2mm}
\subsection{Simulation}
\begin{figure}
    \centering
    \centerline{\includegraphics[width=\linewidth]{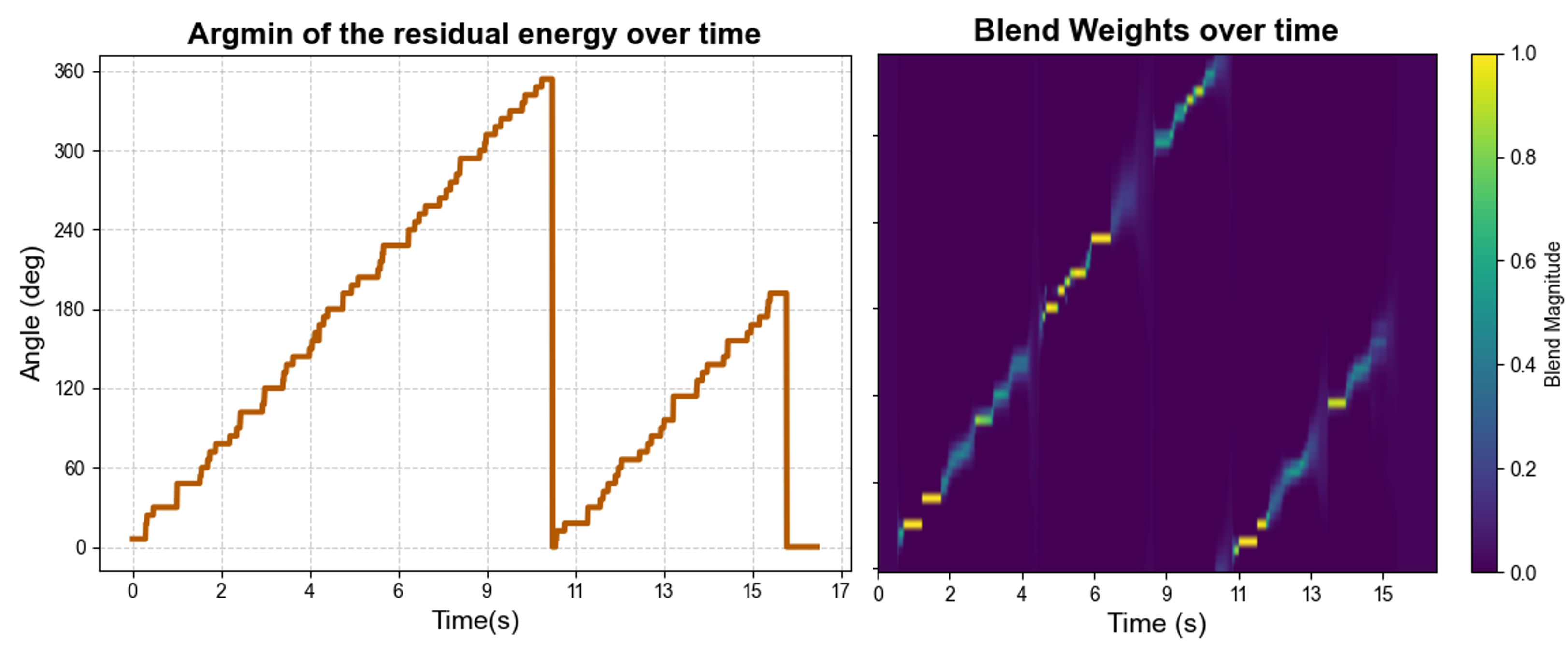}}
    \centerline{\includegraphics[width=\linewidth]{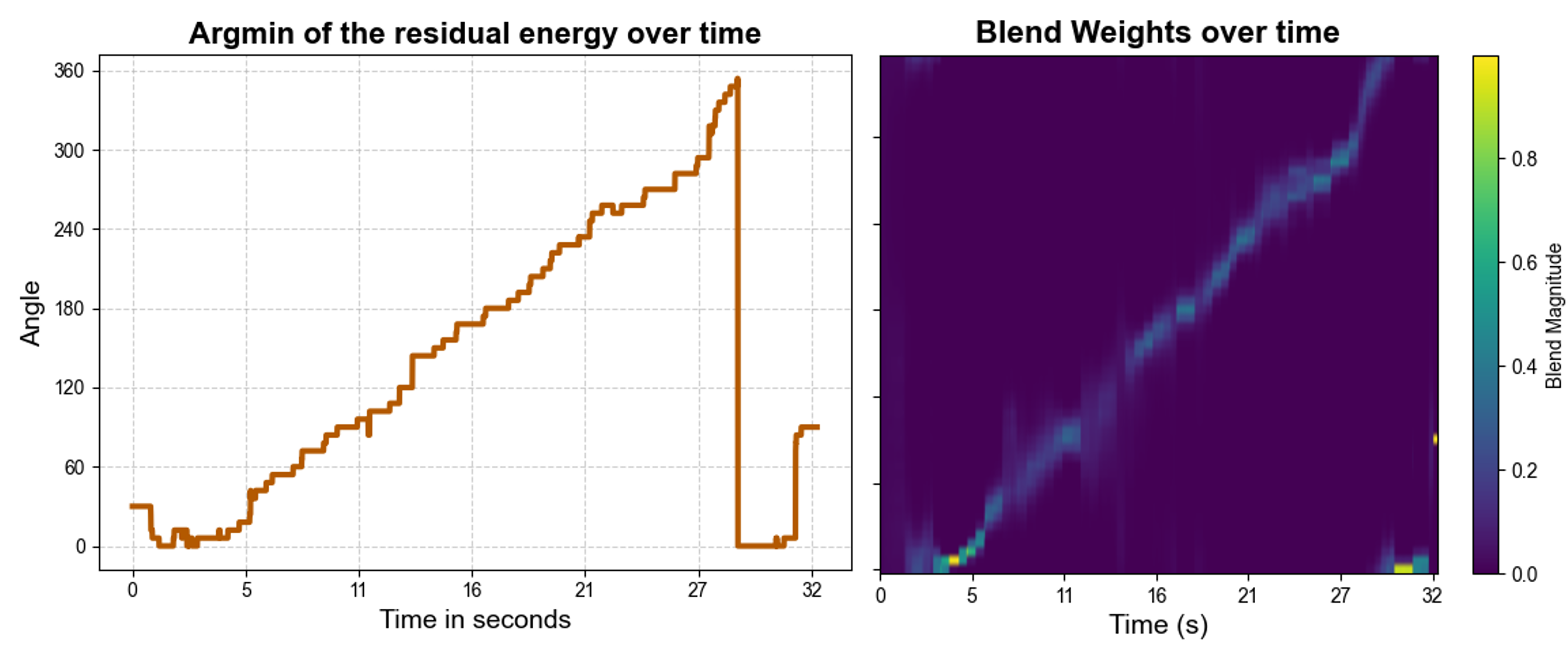}}
    \caption{The argmin of the residual energy at each time and the corresponding blend weights for the grid of beamformers for the simulation (top) and the measurement (bottom).}
    \label{fig:results}
    \vspace{-2mm}
\end{figure}

A continuous motion simulation is performed in pyroomacoustics \cite{scheibler2018pyroomacoustics} within an [8 m, 8 m, 5 m] room (RT60 $\approx$ 200 ms). A 4-microphone array centered at [4 m, 4 m, 2 m] records speech from the EARS dataset \cite{richter2024ears}, sampled at 48 kHz. One talker, initialized at [7 m, 4 m, 2 m] in front of the array, moves in $6^\circ$ azimuth steps, covering each step in 167 ms ($\approx$ 2 m/s). Fig.~\ref{fig:results} (top) shows that the proposed residual-based loss accurately tracks the talker’s motion. The blend weights of the BSM filters further indicate that, under continuous motion, the algorithm adaptively combines filters to render perceptually relevant directions. 
\subsection{Experimental}
\begin{figure}
    \centering
    \centerline{\includegraphics[width=\linewidth]{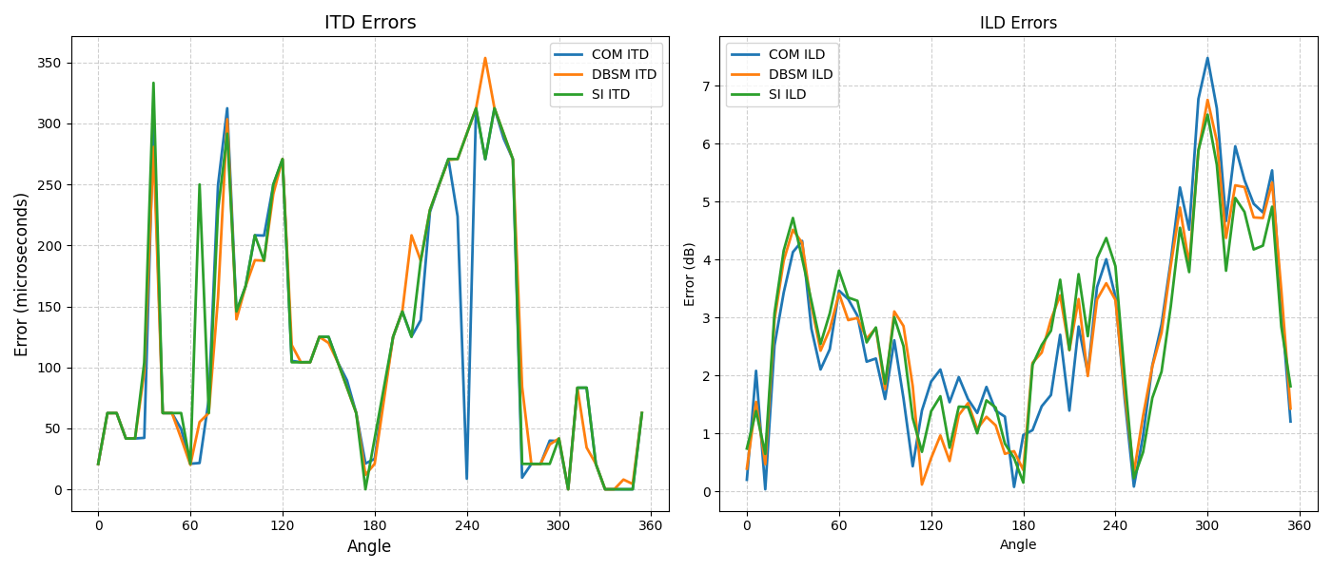}}
    \caption{The error in ITD and ILD over time for the grid of directions considered on average. Results demonstrate that we can manipulate the FoV of the acoustic scene with equally accurate binaural cues as previous methods.}
    \label{fig:interaural}
    \vspace{-2mm}
\end{figure}

The simulation experiment is recreated in a real-world environment with a 4-microphone head-worn microphone array. One talker was asked to start in front of the listener and walk in a counter-clockwise direction, similar to the simulation experiment. Fig.~\ref{fig:results} (bottom) demonstrates the effectiveness of the proposed method in tracking the talker as they walk in the environment. This indicates that, also under real-world conditions, the proposed method is able to render the direct source component with the correct binaural cues as they move through the environment, especially with the MoE module that directly utilizes the HRTFs. Shown in figure ~\ref{fig:interaural} is the error in the ITD and ILD from each of the candidate algorithms. This corresponds to the error is rendering the residual sound field and demonstrates that we are able to match the interaural cues from BSM while providing the user with additional fidelity over the recorded scene. Notably, we could enhance the moving source dynamically over time using the proposed framework with a set of FoVE filters designed to enhance each direction. 

\vspace{-2mm}
\section{Conclusion}
In this work, a novel mixture of experts framework is theorized for binauralization. The proposed framework extends previous work in signal-dependent binauralization to scenarios with continuous motion and for adjustable field-of-view enhancement. Our results demonstrate that the framework is not only effective but highly modular, so that it can be extended to incorporate new advances in the field. For example, future work will focus on utilizing a neural beamspace projection.

\bibliographystyle{IEEEbib}
\bibliography{refs}

\end{document}